0491

# A Feasibility Study of Task-Based fMRI at 0.55 T

Parsa Razmara[1], Takfarinas Medani[1], Anand A. Joshi[1], Majid Abbasi Sisara[1], Ye Tian[1], Sophia X. Cui[2], Justin P. Haldar[1], Krishna S. Nayak[1], and Richard M. Leahy[1]
[1]University of Southern California, Los Angeles, CA, United States, [2]Siemens Healthineers, Los Angeles, CA, United States

## Synopsis

**Motivation:** 0.55T MRI offers advantages compared to conventional field strengths, including reduced susceptibility artifacts and better compatibility with simultaneous EEG recordings. However, reliable task-based fMRI at 0.55T has not been significantly demonstrated.

**Goal(s):** To establish a robust task-based fMRI protocol and analysis pipeline at 0.55T that achieves full brain coverage and results comparable to what is expected for activation extent and location.

**Approach:** We attempted fMRI at 0.55T by combining EPI acquisition with custom analysis techniques. Finger-tapping and visual tasks compared 5 and 10-minute runs to enhance activation detection.

**Results:** Significant activations demonstrating that high-quality task-based fMRI is achievable at 0.55T in single subjects.

**Impact:** This study demonstrates that reliable task-based fMRI is feasible on 0.55T scanners, potentially broadening functional neuroimaging access in clinical and research settings where high-field MRI is unavailable or impractical, supporting broader diagnostic and research applications.

## Introduction

Low-field MRI scanners operating at 0.55 Tesla offer several advantages, such as reduced susceptibility artifacts[1] and better compatibility with simultaneous EEG recordings[2]. However, the spatial coverage limitations and temporal efficiency of this field present challenges to robust task-based fMRI[3]. This study introduces a task-based fMRI protocol and analysis pipeline at a high-performance 0.55T scanner, utilizing a standard single-shot EPI sequence and preprocessing techniques. The imaging protocol achieves full brain coverage and yields activation results in single subjects comparable to what is typically achieved at 3T, demonstrating that high-quality task-based fMRI is achievable at 0.55T.

## Methods

*Data acquisition:* six (one subject excluded) healthy volunteers participated in the finger-tapping task, and six healthy volunteers participated in the visual stimulation task. Imaging was performed on a 0.55T scanner (prototype MAGNETOM Aera, Siemens Healthineers)[4] with high-performance gradients using a 16-channel head-neck receive array.

Anatomical images were acquired using a T1-weighted MPRAGE sequence with a 1 mm isotropic resolution. Functional data were acquired with a single-shot EPI BOLD sequence (FOV: 210 × 170 mm$^2$, voxel size: 3.3 × 3.3 × 4 mm$^3$, slice gap: 0.8 mm, TR: 3000 ms, TE: 85 ms), balancing temporal resolution and coverage[5-10]. Twenty-one interleaved axial slices covered the entire brain.

*Experiments and Data Analysis:* The finger-tapping task employed a block design with alternating 30-second blocks of self-paced bilateral finger-tapping and rest, totaling 5 minutes. The visual stimulation task involved a flashing checkerboard, presented in 24-second blocks alternating with rest, over a ~5-minute scan. During the rest period, a cross (+) fixation point was presented on the screen to minimize eye movements.

Data preprocessing utilized the BrainSuite fMRI Pipeline (BFP)[11,12] and Statistical Parametric Mapping (SPM12)[13], including slice timing correction, motion correction, spatial smoothing with an 8 mm FWHM, and high-pass temporal filtering at 0.005 Hz. Functional images were co-registered to individual T1 images and normalized to the USCBrain atlas[14] (MNI for SPM) for accurate anatomical localization.

Activation maps were generated using a voxel-wise GLM. The expected BOLD response was modeled by convolving the task design with a canonical HRF[13]. Statistical significance was assessed using t-tests, and results were thresholded using FDR correction (q < 0.05). To assess the impact of extended scan durations, two analyses were performed: first, two consecutive runs were concatenated to double the time points; second, corresponding time points from the two runs were averaged.

## Results

Robust activation was observed in task-relevant cortical areas across all participants. In the finger-tapping task, significant activation (p < 0.05, FDR corrected) was detected bilaterally in the primary motor cortex (M1) and supplementary motor area (SMA). In the visual stimulation task, significant activation (Fig. 1) was observed in the primary visual cortex (V1) and adjacent visual areas. Average t-values in these regions increased with extended scan durations, indicating enhanced statistical power (Fig 2, Fig 3). The concatenated data showed improved correlation stability and reduced variability in non-target regions compared to single-run analyses (Fig. 4).

Averaging data timepoints led to a significant increase in correlation coefficients within the ROIs, increasing to approximately 0.75 (Fig. 5).

## Discussion

This study demonstrates that robust task-based fMRI is achievable at 0.55T, supporting the feasibility suggested by initial studies[3]. We utilized an EPI BOLD protocol with full brain coverage with an in-plane resolution of 3.3 mm and slice thickness of 4 mm.

Our imaging and analysis pipeline yielded statistical metrics comparable to those typically observed in 3T studies. For the finger-tapping task, we achieved a mean t-score of 12.67 ($p<0.05$), aligning with typical 3T results around t~10 in motor cortex activations[16,17]. For the visual task, our mean t-score was 8 ($p<0.05$), consistent with standard 3T activations in the primary visual cortex[15,18]. These values indicate significant task-related activations, demonstrating that low-field 0.55T fMRI can achieve robust, statistically comparable results to 3T systems with extended scan durations, thereby addressing the lower SNR and BOLD sensitivity associated with low-field fMRI.

Compared to previous 0.55T studies, our approach not only provides full brain coverage but also enhances activation statistics, achieving results comparable to conventional field systems. The ability to achieve such results on a 0.55T scanner underscores the potential of low-field fMRI for functional neuroimaging applications.

## Conclusion

This study demonstrates the feasibility of task-based fMRI at 0.55T with full brain coverage using an EPI BOLD sequence and analysis pipeline. The robust activations and high correlation values achieved are comparable to higher field strengths, supporting the viability of low-field scanners for functional neuroimaging. Since 0.55T is better compatible with simultaneous EEG recordings, it can be used to map the brain in multimodal modes, specifically EEG-fMRI[2].

## Acknowledgements

The authors receive grant support from the National Institutes of Health (R01EB026299, U01EB023820, R21-HL159533, R01-AR078912, U01-HL167613), National Science Foundation (No. 1828736) and receive research support from Siemens Healthineers.

# Figures

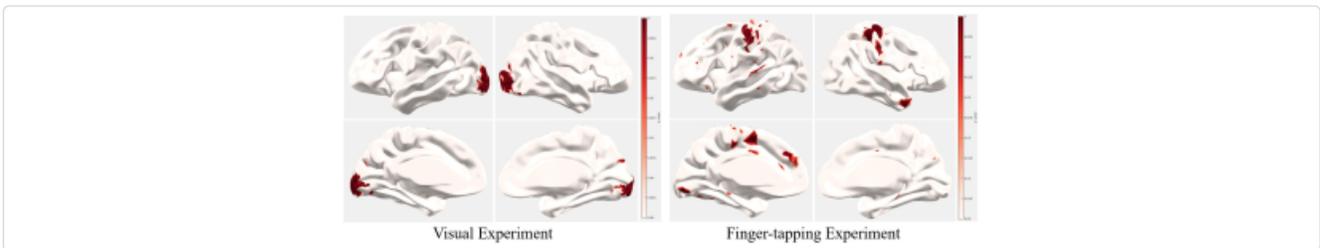

**Figure 1:** Unthresholded p-value maps of task-based fMRI activation in a single subject during 5-minute visual (left) and finger-tapping (right) tasks, corrected using FDR. The 3D brain maps display p-values with a reversed colormap ranging from 0.05 to 0, where values above 0.05 are at the upper limit and 0 marks the most significant activations. These unthresholded maps without clustering can reveal the actual behavior, highlighting task-related activations in target ROIs and adjacent areas.

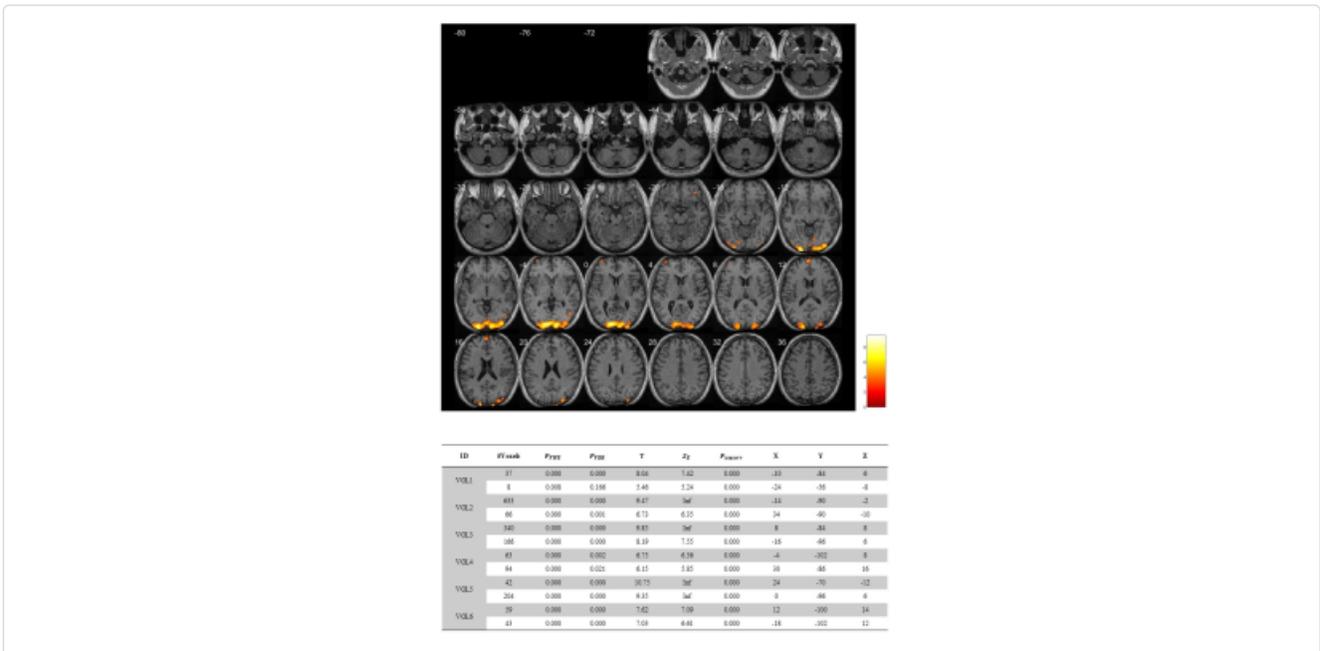

**Figure 2: Activation maps and statistical metrics for the visual experiment.** For each subject, two activated clusters from concatenated 10-minute runs are presented, displaying the number of voxels in each cluster, p-values, T-scores, and Z-scores to assess activation reliability and intensity. The figure shows a single subject's brain activation map overlaid on anatomical slices, emphasizing areas of significant activation in response to the visual task, illustrating task-specific cortical engagement.

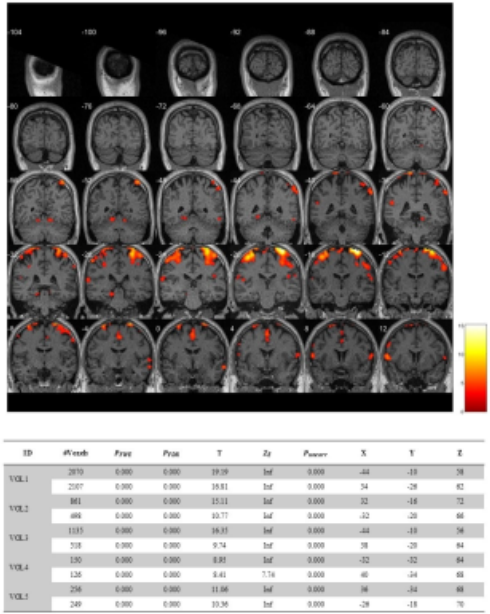

Figure 3: Statistical maps and values for the bilateral finger-tapping task. Each subject's left and right hemisphere clusters from concatenated 10-minute runs are presented, displaying voxel counts, p-values, T-scores, and Z-scores to assess activation reliability and intensity. The single subject's brain activation map overlaid on anatomical slices, highlights significant motor cortex activation in response to bilateral finger-tapping, with statistical metrics detailing cluster size and robustness.

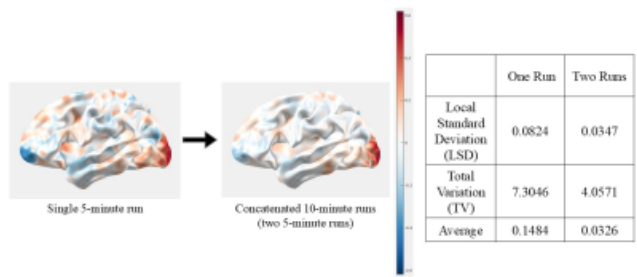

Figure 4: Effect of Longer Scan Times on Activation Robustness at 0.55 T for the Visual Experiment. Extended scan duration (concatenated runs) enhances statistical power, yielding focused activation in target regions and reducing spurious activation in non-target areas, thus improving specificity and robustness. The table shows average values in three non-target ROIs (each with 200 voxels outside the visual cortex), with lower Local Standard Deviation (LSD) and Total Variation (TV) indicating increased homogeneity and consistency across the cortex.

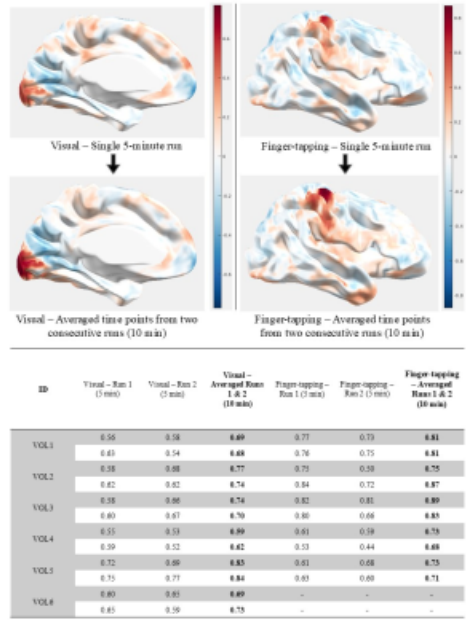

**Figure 5: Effect of Averaging on Activation Robustness in a Single Subject.** Averaging time points from two consecutive runs enhances correlation coefficients within the ROIs. This method reduces transient noise and external variability, yielding stronger, more reliable activation patterns and improving the robustness of detected activations compared to a single run. The table compares peak correlation values across all volunteers, highlighting increased consistency with extended scan duration.